\documentclass[preprint,amsmath,showpacs]{revtex4}
\usepackage{graphicx}% Include figure files
\usepackage{dcolumn}% Align table columns on decimal point
\usepackage{bm}% bold math

\def\bee{\begin{eqnarray}}
\def\eee{\end{eqnarray}}

\begin{document}
\draft
\title{Probing the octant of $\theta_{23}$ with very long baseline neutrino oscillation experiments: a global look  }
\author{Guey-Lin Lin$^{a,b}$\footnote{E-mail: glin@cc.nctu.edu.tw}
and Yoshiaki Umeda$^a$\footnote{E-mail: umeda@faculty.nctu.edu.tw} }
\address{
$^a$Institute of Physics, National Chiao-Tung University, Hsinchu
300,
Taiwan \\
$^b$Physics Division, National Center for Theoretical Sciences,
Hsinchu 300, Taiwan }
\date{\today}
\begin{abstract}
We investigate the baseline range in which the $\theta_{23}$
degeneracy in neutrino oscillation probabilities is absent for fixed
values of $\theta_{13}$ and CP violation phase $\delta_{\rm CP}$. We
begin by studying sensitivities of neutrino oscillation
probabilities to $\theta_{13}$, $\theta_{23}$ and $\delta_{\rm CP}$
for very-long-baseline neutrino oscillations. We show contour graphs
of the muon-neutrino survival probability $P(\nu_{\mu}\to
\nu_{\mu})$ and the appearance probability $P(\nu_e\to \nu_{\mu})$
on the $\cos 2\theta_{23}-\sin 2\theta_{13}$ plane for baseline
lengths $L=1000, \ 5000, \ 10000$, and $12000$ km.  For each
baseline length, it is found that $P(\nu_{\mu}\to \nu_{\mu})$ is
more sensitive to $\sin 2\theta_{13}$ at energies around its local
maximum while it is more sensitive to $\cos 2\theta_{23}$ at
energies around its local minimum. On the other hand, the appearance
probability $P(\nu_e\to \nu_{\mu})$ is sensitive to
$\sin2\theta_{13}$ and $\cos2\theta_{23}$ only near its local
maximum. We observe that the $\theta_{23}$ degeneracy in
$P(\nu_{\mu}\to \nu_{\mu})$ is absent at energies around the local
maximum of this probability, provided $\theta_{13}$ is sufficiently
large. The $\theta_{23}$ degeneracy is also absent in general near
the local maximum of $P(\nu_e\to \nu_{\mu})$. Using analytic
approximations for neutrino oscillation probabilities, we
demonstrate that the above observations for $L=1000, \ 5000, \
10000, \ {\rm and} \ 12000$ km are in fact valid for all distances.
The implications of these results on probing the octant of
$\theta_{23}$ are discussed in details.
\end{abstract}
\pacs{14.60.Pq, 13.15.+g, 14.60.Lm }
\maketitle
\section{Introduction}
The understanding of neutrino masses and mixing matrix is crucial to
unveil the mystery of lepton flavor structures. The updated SK
analysis of the atmospheric neutrino data gives \cite{Ashie:2005ik}
\begin{equation}
 1.5\cdot 10^{-3}\,\,\, {\rm eV}^{2}< \vert \Delta m_{31}^{2}\vert < 3.4\cdot
 10^{-3}\,\,\, {\rm eV}^{2},  \ \sin^{2}2\theta_{23} >0.92. \label{range23_new}
\end{equation}
This is a  $90\% \, {\rm C.L.}$ range with the best fit values given
by $\sin^{2}2\theta_{23} =1$ and $\Delta m_{31}^{2}=2.1\cdot
10^{-3}\, \, \, {\rm eV}^{2}$ respectively. An earlier result based
upon $L/E$ analysis gives \cite{Ashie:2004mr}
\begin{equation}
 1.9\cdot 10^{-3}\,\,\, {\rm eV}^{2}< \vert \Delta m_{31}^{2}\vert < 3.0\cdot
 10^{-3}\,\,\, {\rm eV}^{2},  \ \sin^{2}2\theta_{23} >0.9. \label{range23}
\end{equation}
at $90\% \, {\rm C.L.}$ where the best fit values are given by
$\sin^{2}2\theta_{23} =1$ and $\Delta m_{31}^{2}=2.4\cdot 10^{-3}\,
\, \, {\rm eV}^{2}$ respectively. The scenario of $\nu_{\mu}\to
\nu_{\tau}$ oscillation for atmospheric neutrinos has been confirmed
by the K2K experiment \cite{Aliu:2004sq,Ahn:2006zz}. Furthermore the
results in the solar neutrino oscillation measurements are also
confirmed by KamLAND reactor measurements
\cite{Eguchi:2002dm,Araki:2004mb}. Combining these measurements, the
LMA solution of the solar neutrino problem is established and the
updated $2\sigma$ parameter ranges are given by \cite{Fogli:2005cq}
\begin{equation}
7.21\cdot 10^{-5}\,\,\, {\rm eV}^{2}< \Delta m_{21}^{2}< 8.63\cdot
10^{-5}\, \, \, {\rm eV}^{2}, \ 0.267< \sin^{2}\theta_{12}< 0.371,
\label{range12}
\end{equation}
with the best fit values $\Delta m_{21}^{2}=7.92\cdot 10^{-5}\, \,
\, {\rm eV}^{2}$ and $\sin^{2}\theta_{12}=0.314$.

Despite the achievements so far in measuring the neutrino mixing
parameters, the sign of $\Delta m_{31}^2$, the mixing angle
$\theta_{13}$ and the CP violating parameter $\delta_{\rm CP}$ in
the mixing matrix remain to be determined. Furthermore, one is keen
to resolve the octant degeneracy of $\theta_{23}$
\cite{Fogli:1996pv}.

The mixing angle $\theta_{13}$ is constrained by the reactor
experiments \cite{Apollonio:1999ae,Boehm:2001ik}. The CHOOZ
experiment \cite{Apollonio:1999ae} gives a more stringent constraint
on $\theta_{13}$ with $\sin^2 2\theta_{13} < 0.1$ for a large
$\Delta m_{31}^2$ (90\% C.L.). A recent global fit based upon
three-flavor neutrino oscillation gives the $2\sigma$ upper bound,
$\sin^2 2\theta_{13}< 0.124$ \cite{Fogli:2005cq}. It is well known
that the mixing angle $\theta_{13}$ can be enhanced by the matter
effect in Earth. The appearance oscillations $\nu_{\mu}\to \nu_e$,
$\nu_e\to \nu_{\mu}$, and the survival mode $\nu_{\mu}\to \nu_{\mu}$
performed in a very-long baseline have been proposed
\cite{Mocioiu:2000st} to probe the angle $\theta_{13}$ and the sign
of $\Delta m_{31}^2$. Furthermore, the aforementioned very long
baseline neutrino experiments as well as future atmospheric neutrino
experiments are proposed to determine the deviation of $\theta_{23}$
to maximality \cite{long_theta23}. In this work, we focus on the
mixing angle $\theta_{23}$. We shall provide a global survey on
ideal neutrino energies in the GeV range and baseline lengths from
$10^3$ km to $10^4$ km for probing the octant of the mixing angle
$\theta_{23}$. The muon neutrino survival probability
$P(\nu_{\mu}\to \nu_{\mu})\equiv P_{\mu\mu}$ and electron neutrino
appearance probability $P(\nu_e\to \nu_{\mu})\equiv P_{e\mu}$ are
both studied for this purpose. We observe that the muon neutrino
survival probability $P_{\mu\mu}$ has complementary dependencies on
mixing angles $\theta_{13}$ and $\theta_{23}$ as the neutrino energy
varies. This property is established by studying the dependencies of
$P_{\mu\mu}$ on $\cos 2\theta_{23}$ and $\sin 2\theta_{13}$ while
keeping other parameters fixed. The choice of the parameter $\cos
2\theta_{23}$ is appropriate as
\begin{equation}
\frac{1}{2}\cos2\theta_{23}=\frac{1}{2}-\sin^2\theta_{23},
\end{equation}
which is a probe to the deviation of $\theta_{23}$ to the best-fit
value $\pi/4$. We find that the dependencies of $P_{\mu\mu}$ on
$\cos 2\theta_{23}$ and $\sin 2\theta_{13}$ at energies near local
maxima of this probability differ drastically from those at energies
near local maxima of the same probability. In the former case, the
probability $P_{\mu\mu}$ is always more sensitive to $\sin
2\theta_{13}$. Furthermore, the $\theta_{23}$ degeneracy is absent
in this case. In the latter case, the probability $P_{\mu\mu}$ is
more sensitive to $\cos 2\theta_{23}$ while the $\theta_{23}$
degeneracy is generally present. Such information is useful for
probing the octant of $\theta_{23}$. We also study sensitivities of
the probability $P_{e\mu}$ to $\cos 2\theta_{23}$ and $\sin
2\theta_{13}$ with other parameters fixed. We only focus on energies
near the local maximum of $P_{e\mu}$ as this probability is not
sensitive to mixing parameters for energies near its local minimum.

This paper is organized as follows. In Section II, we compare results
on the oscillation probability $P_{e\mu}$ obtained by the full
calculation with those obtained by various analytic approximations.
This comparison is essential since analytic approximations will be
employed for discussions in later sessions. To set up the analytic
approximation, we introduce the concept of average density which
varies with the total neutrino path-length inside the Earth.
Applying full calculations and the two-layer analytic approximations
\cite{Freund:1999vc}, we identify the energy values for local maxima
and local minima of neutrino oscillation probabilities $P_{\mu\mu}$
and $P_{e\mu}$ for baseline lengths $1000 \leq L/{\rm km}\leq
12000$. It is found that the two-layer approximation is quite
satisfactory compared to the full calculation for computing these
energy values. In Section III, we first present the dependencies of
$P_{\mu\mu}$ and $P_{e\mu}$ on the CP violation phase $\delta_{\rm
CP}$. It will be shown that, unlike $P_{e\mu}$, $P_{\mu\mu}$ is not
sensitive to the CP violation phase $\delta_{\rm CP}$. We study
numerically the effect of CP violation phase to the appearance
probability $P_{e\mu}$. The result confirms the so-called magic
baseline \cite{Barger:2001yr,Huber:2003ak,Smirnov:2006sm} at
$L\approx 7600$ km where $P_{e\mu}$ is rather insensitive to the CP
violation phase. After discussions on the CP violation phase, we
present the contour graphs of probabilities $P_{\mu\mu}$ and
$P_{e\mu}$ on $\cos 2\theta_{23}-\sin 2\theta_{13}$ plane for
baseline lengths $L=1000, \ 5000, \ 10000$, and $12000$ km. At all
these baseline lengths, we shall see that $P_{\mu\mu}$ is more
sensitive to $\sin 2\theta_{13}$ at energies around its local
maximum while it is more sensitive to $\cos 2\theta_{23}$ at
energies around its local minimum. Such observations are then  justified by 
using the two-layer
analytic approximations for neutrino oscillation probabilities. With
this approximation, the baseline lengths and neutrino energies
allowing an unambiguous determination of $\theta_{23}$ through
measuring $P_{\mu\mu}$ are identified.  In Section IV, we discuss the prospects of probing the
$\theta_{23}$ octant via measuring $P_{e\mu}$ and $P_{\mu\mu}$.
We then conclude in the same section.
\section{The comparison of full calculations and analytic approximations }

We begin the discussions with the relation connecting flavor and
mass eigenstates of neutrinos, $\nu_{\alpha}=\sum_i U_{\alpha
i}\nu_{i}$, with $U$ the Maki-Nakagawa-Sakata mixing matrix
\cite{MNS} given by
\begin{equation}
U =\left(\!
\begin{array}{ccc}
c_{12}c_{13} & s_{12}c_{13} & s_{13}e^{-i\delta_{\rm CP}} \\
-s_{12}c_{23}-c_{12}s_{13}s_{23}e^{i\delta_{\rm CP}} &
c_{12}c_{23}-s_{12}s_{13}s_{23}e^{i\delta_{\rm CP}}
& c_{13}s_{23} \\
s_{12}s_{23}-c_{12}s_{13}c_{23}e^{i\delta_{\rm CP}} &
-c_{12}s_{23}-s_{12}s_{13}c_{23}e^{i\delta_{\rm CP}} & c_{13}c_{23}
\end{array}
\!\right)\,,
\end{equation}
where $s_{ij}$ and $c_{ij}$ denote $\sin\theta_{ij}$ and
$\cos\theta_{ij}$, respectively. The value for the Dirac type
CP-phase $\delta_{\rm CP}$ ranges from 0 to 2$\pi$. The evolutions
of neutrino flavor eigenstates are governed by the equation
\begin{eqnarray}
i\frac{d}{dt} |\nu(t)\rangle &=&  \left\{ \frac{1}{2E_\nu}U \left(\!
\begin{array}{ccc}
0 & 0               & 0 \\
0 & \Delta m^2_{21} & 0 \\
0 & 0               & \Delta m^2_{31}
\end{array}
\right)U^\dagger  + \left(\!
\begin{array}{ccc}
V & 0 & 0 \\
0 & 0 & 0 \\
0 & 0 & 0
\end{array}
\!\right)\right\} |\nu(t)\rangle \,, \label{evolution}
\end{eqnarray}
where $|\nu(t)\rangle = (\nu_e(t), \nu_\mu(t), \nu_\tau(t))^T$,
$\Delta m^2_{ij} \equiv m^2_{i}-m^2_{j}$ is the mass-squared
difference between the $i$-th and $j$-th mass eigenstates, and
$V\equiv \sqrt{2}G_F N_e$ is the effegtive potential arising from
the charged current interaction between $\nu_e$ and electrons in the
medium with $N_e$ the electron number density. Numerically $V= 7.56
\times 10^{-14}$ $(\rho/[{\rm g/cm^3}])$$Y_e$ [{\rm eV}] with $Y_e$
denoting the number of electrons per nucleon. We take $Y_e\sim 0.5$
in our calculations. One solves Eq.~(\ref{evolution}) by
diagonalizing the Hamiltonian on its right hand side. This amounts
to writing the right hand side of Eq.~(\ref{evolution}) as
$U'H'U^{'\dagger}|\nu(t)\rangle$ with $U'$ the neutrino mixing
matrix in the matter and $H' \equiv {\rm diag}(E_1, E_2, E_3)$ the
Hamiltonian after diagonalization. To obtain various oscillation
probabilities described later, we have used the parametrization in
\cite{profile} for the Earth density profile.

\begin{figure}[tb]
\begin{center}
$\begin{array}{c}
\includegraphics[width=9.5cm]{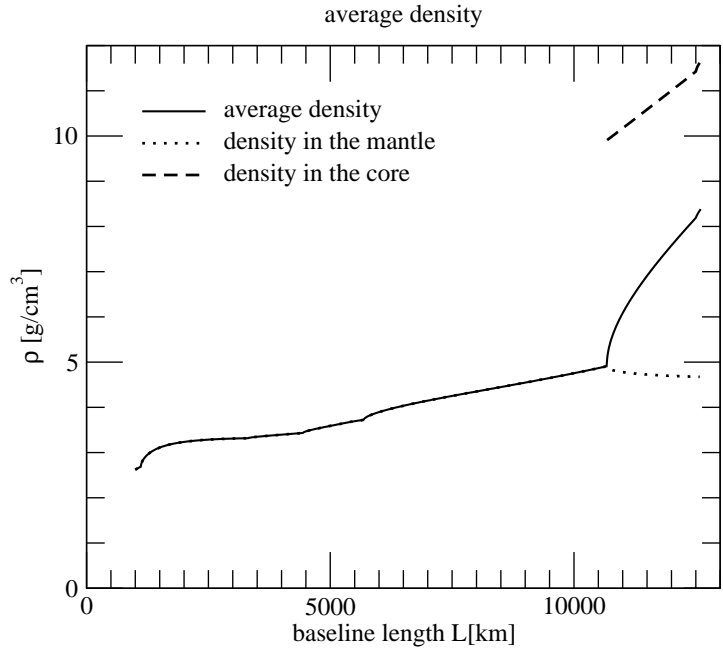}
\end{array}$
\end{center}
\caption{The average Earth density along the path traversed by the
neutrino as a function of the path length $L$.} \label{Fdensity}
\end{figure}
For analytic calculations, we employ the two-layer approximation for
the Earth density profile \cite{Freund:1999vc}. Given a path-length
$L$ for a neutrino traversing the Earth medium, one can divide $L$
into the sum $L=L_1+L_2+\cdots L_n$ with each $L_i$ corresponding to
a region with a specific matter density. The average density for
this path-length is then given by $\rho=(\rho_1L_1+\rho_2L_2+\cdots
\rho_n L_n)/L$. The Earth medium can be categorized as the Earth
mantle and the Earth core. If a neutrino only traverses the Earth
mantle, we shall use the one-density approximation for the analytic
calculation with the density defined by the above prescription.
However, if a neutrino traverses both the Earth mantle and the Earth
core, one should write the total neutrino path-length as
$L=2L_m+L_c$ with
\begin{eqnarray}
L_m&=&R\left(\cos\theta_n-\sqrt{\frac{r_c^2}{R^2}-\sin^2\theta_n}\right),\nonumber
\\
L_c&=&2R\sqrt{\frac{r_c^2}{R^2}-\sin^2\theta_n}, \label{core_mantle}
\end{eqnarray}
where $R=6371$ km and $r_c=3480$ km are the radii of the entire
Earth and the Earth core respectively while $\theta_n$ is the
incident Nadir angle of the neutrino. We note that the critical
Nadir angle for a neutrino to pass the Earth core is $33.17^{\circ}$
corresponding to $L=10674$ km. For $L> 10674$ km, one separately
defines average densities within the path-length $L_m$ and the
path-length $L_c$ respectively. The average densities as functions
of the neutrino path-length is shown in Fig.~\ref{Fdensity}. For
$L\leq 10674$ km, there is only one curve for the average density,
which is represented by the solid line in the figure. Beyond this
distance, one can define the average density in the core and the
average density in the mantle, which are represented by dashed and
dotted lines respectively. Alternatively, one can also define single
average density for $L> 10674$ km by ignoring the distinction
between the mantle and the core. This is seen from the solid line
for $L>10674$ km. However, in our analytic calculations, we shall
adopt the two-density approach for $L>10674$ km.
\begin{figure}[tb]
\begin{center}
$\begin{array}{c}
\includegraphics[width=9.5cm]{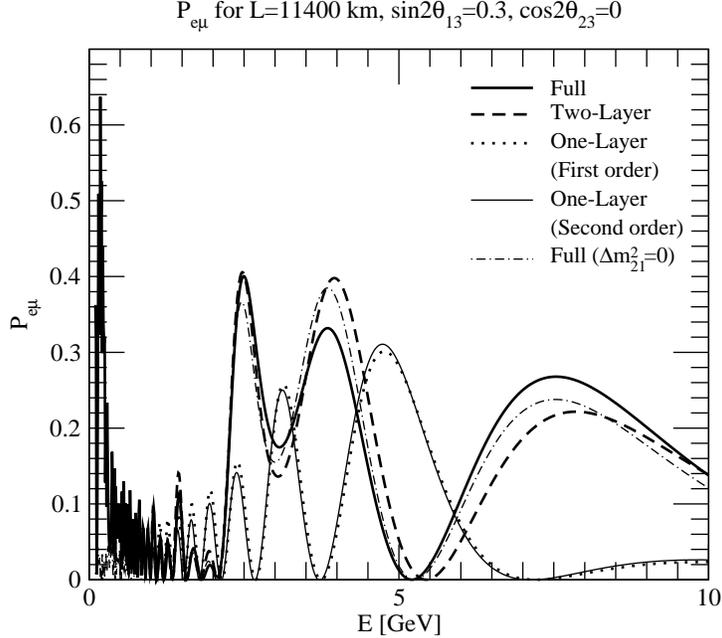}
\end{array}$
\end{center}
\caption{A comparison of $P_{e\mu}$ obtained by the full numerical
calculation and various approximations. The thick solid curve
denotes the result by the full numerical calculation. The
dotted-dashed curve denotes the result by setting $\Delta
m_{21}^2=0$ in the full numerical calculation. The dashed curve
represents the result obtained by the two-layer approximation in the
leading order of $\tau\equiv \Delta m_{21}^2/\Delta m_{31}^2$. The
dotted curve denotes the result obtained by one-density
approximation in the leading order of $\tau$ while the thin solid
curve is that obtained by the one-density approximation in the
next-to-leading order of $\tau$. } \label{FpmmeL11400}
\end{figure}

For analytic calculations, we only compute oscillation probabilities
up to the lowest order in $\tau\equiv \Delta m_{21}^2/\Delta
m_{31}^2$. In other words, we set $\Delta m_{21}^2$=0 in analytic
calculations and consequently the mixing angle $\theta_{12}$ and the
CP phase $\delta_{\rm CP}$ drop out from the oscillation
probabilities. The probabilities $P_{\mu\mu}$ and $P_{e\mu}$ in the
two-layer approximations are given
by\cite{Chizhov:1999he,Akhmedov:1998ui,Petcov:1987cd,Bernabeu:2001xn,Hsu_thesis}:
\begin{eqnarray}
P_{\mu\mu}&=&\cos^4 \theta_{23}+\left(u^2+v^2\right)\sin^4
\theta_{23}+2\cos^2\theta_{23}\sin^2\theta_{23}\left(u\cos t+v\sin
t\right),\nonumber \\
P_{e\mu}&=&\sin^2 \theta_{23}\left(1-u^2-v^2\right).\label{analytic}
\end{eqnarray}
The quantities $u$, $v$ and $t$ are defined as
\begin{eqnarray}
u&=&\cos(2\phi^m)\cos(\phi^c)-\cos(2\theta_{13}^c-2\theta_{13}^m)\sin(2\phi^m)\sin(\phi^c),\nonumber
\\
v&=&-\cos(2\theta_{13}^m)\left[\sin(\phi^c)\cos(2\phi^m)\cos(2\theta_{13}^c-2\theta_{13}^m)+\cos(\phi^c)\sin(2\phi^m)
\right]\nonumber \\
&&+\sin(2\theta_{13}^m)\sin(\phi^c)\sin(2\theta_{13}^c-2\theta_{13}^m),\nonumber
\\
 t&=&\frac{(M_{13}^2)^m+(m_{13}^2)^m}{4E}\times
2L^m+\frac{(M_{13}^2)^c+(m_{13}^2)^c}{4E}\times L^c,
\end{eqnarray}
where
\begin{eqnarray}
\phi^{m(c)}&=&\frac{\Delta_{31}^{m(c)}}{4E}L^{m(c)},\nonumber \\
(M_{13}^2)^{m(c)}&=&(\Delta
m_{31}^2+A_e^{m(c)}+\Delta_{31}^{m(c)})/2,\nonumber \\
(m_{13}^2)^{m(c)}&=&(\Delta
m_{31}^2+A_e^{m(c)}-\Delta_{31}^{m(c)})/2,
\end{eqnarray}
with
\begin{equation}
\Delta_{31}^{m(c)}=\sqrt{(\Delta m_{31}^2\sin
2\theta_{13})^2+(A_e^{m(c)}-\Delta m_{31}^2\cos 2\theta_{13})^2}.
\end{equation}
 The superscripts $m$ and $c$ denote quantities defined in the
Earth mantle and the Earth core respectively. For neutrinos
traversing only the Earth mantle, one simply sets $L_c=0, \ 2L_m=L$
in the above equations and recovers well known expressions for
$P_{\mu\mu}$ and $P_{e\mu}$ in the one-density approximation
\cite{Akhmedov:2004ny}.

\begin{figure}[tb]
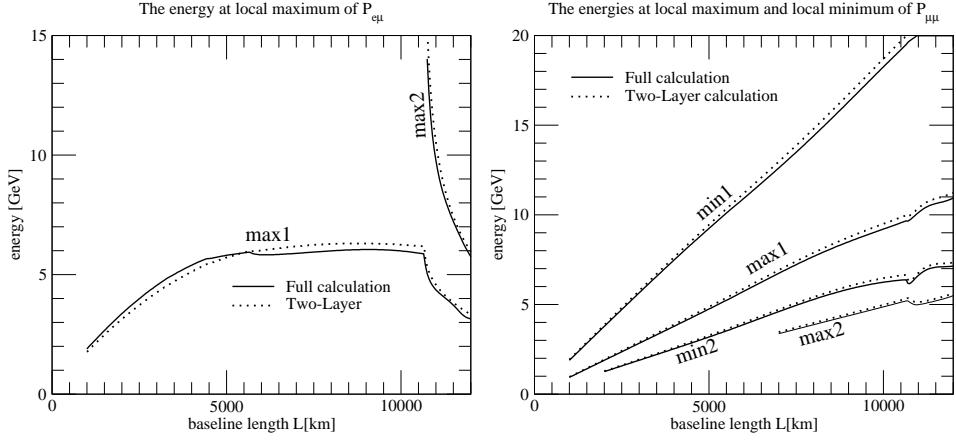

\begin{center}
$\begin{array}{cc}
\includegraphics*[width=6.2cm]{Fpmme_peak.eps} &
\includegraphics*[width=6.2cm]{Fpmmm_peak.eps}
\end{array}$
\end{center}
\caption{Left panel: the energy at the local maximum of $P_{e\mu}$,
as a function of $L$. Right panel: energies at local maxima and
local minima of $P_{\mu\mu}$, as functions of $L$.  } \label{maxmin}
\end{figure}
The accuracy of two-layer approximation is shown in
Fig.~\ref{FpmmeL11400} with a comparison of this approximation to
the full numerical calculation and other approximations. In the
calculations, we have assumed the normal mass hierarchy and taken
$\sin2\theta_{13}=0.3$, $\cos2\theta_{23}=0$, $\delta_{\rm CP}=0$,
$\Delta m_{31}^2=2.4\times 10^{-3}$ eV$^2$, $\Delta
m_{21}^2=8.2\times 10^{-5}$ eV$^2$, and $\tan^2\theta_{12}=0.39$
\cite{Bahcall:2004ut}. This set of parameters will be adopted for
later calculations unless specific mentioning of other choices. This
set of parameters differ from the most updated best-fit values
quoted right after Eq.~(\ref{range12}). However, both set of
parameters give undistinguishable results on $P_{e\mu}$ and
$P_{\mu\mu}$ in the energy range concerned here. A comparison made
at $L=11400$ km has two purposes. First of all, it is known that the
series expansion in the parameter $\tau$ is valid for $L/E_{\nu}\ll
10^4$ (km/GeV) \cite{Akhmedov:2004ny,Mocioiu:2001jy}. Hence analytic
calculations performed at this baseline length test the marginal
region of the condition $L/E_{\nu}\ll 10^4$ (km/GeV). Secondly this
path-length implies that the neutrino traverses both the Earth
mantle and the Earth core. Therefore it is also a good test to the
two-layer approximation. It is seen that the two-layer
approximation, unlike the one-layer approximation, reproduces well
the peak energies of $P_{e\mu}$, while it gives peak probabilities
deviating from those obtained from the full calculation by
$15\%-20\%$. We also see that the two-layer approximation agrees
well with the full calculation in the limit $\Delta m_{21}^2=0$.

For later analysis, we compute energies at local maxima of
$P_{e\mu}$ and those at local maxima and local minima of
$P_{\mu\mu}$ for the baseline range $1000\leq L/{\rm km}\leq 12000$.
The results are depicted in Fig.~\ref{maxmin}. We do not study local
minima of $P_{e\mu}$ since their values are not sensitive to mixing
angles $\theta_{13}$ and $\theta_{23}$. It is seen that the analytic
approximation is satisfactory for computing energies at local maxima
and local minima of neutrino oscillation probabilities. We point out
that the energy curves in Fig.~\ref{maxmin} are calculated with
$\sin2\theta_{13}=0.3$ and $\cos2\theta_{23}=0$. It is found that
these curves are not sensitive to the values of $\sin2\theta_{13}$
and $\cos2\theta_{23}$.

\section{Conditions for the absence of $\theta_{23}$ degeneracy in $P_{\mu\mu}$ and $P_{e\mu}$ at different energies}

\subsection{The dependencies of $P_{\mu\mu}$ and $P_{e\mu}$ on $\delta_{\rm CP}$}
\begin{figure}[tb]
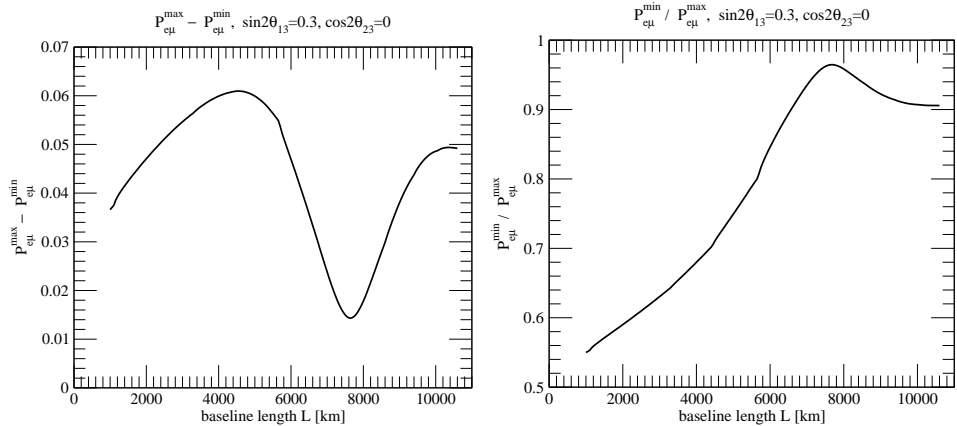

\begin{center}
$\begin{array}{cc}
\includegraphics*[width=6.2cm]{FPem_diff.eps} &
\includegraphics*[width=6.2cm]{FPem_rat.eps}
\end{array}$
\end{center}
\caption{The difference and the ratio of $P_{e\mu}^{\rm max}$ and
$P_{e\mu}^{\rm min}$ as functions of the baseline length.}
\label{emu-cp}
\end{figure}

Before concentrating on $\theta_{13}$ and $\theta_{23}$ dependencies
of neutrino oscillation probabilities, we first study the CP phase
dependencies with the full numerical calculations.
\begin{figure}[tb]
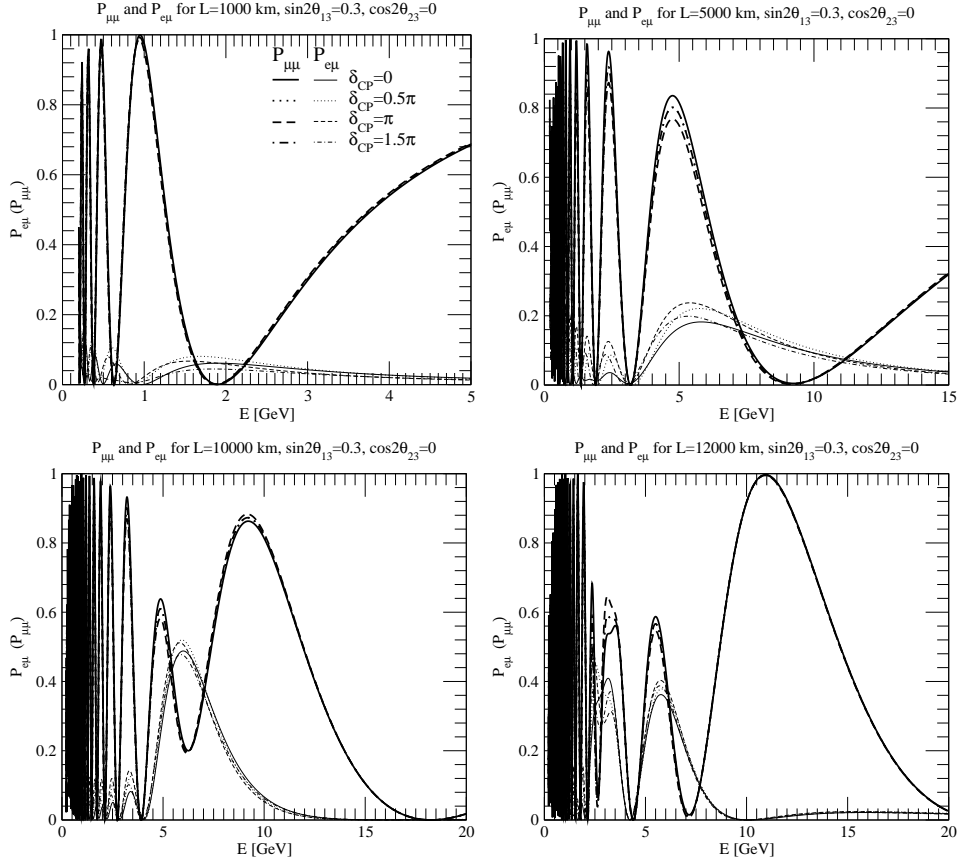

\begin{center}
$\begin{array}{cc}
\includegraphics*[width=6.2cm]{FL1000cp.eps} &
\includegraphics*[width=6.2cm]{FL5000cp.eps}
\\
\includegraphics*[width=6.2cm]{FL10000cp.eps} &
\includegraphics*[width=6.2cm]{FL12000cp.eps}
\end{array}$
\end{center}
\caption{The CP phase dependencies of $P_{\mu\mu}$ and $P_{e\mu}$
for $L=1000$ km, $5000$ km, $10000$ km and $12000$ km respectively.
The probability $P_{\mu\mu}$ is described by the thick curve while
$P_{e\mu}$ is described by the thin curve. }
\label{cp-phase}
\end{figure}
It is easily seen from Fig.~\ref{cp-phase} that $P_{\mu\mu}$ is not
sensitive to the CP phase for all distances displayed. On the other
hand, $P_{e\mu}$ is rather sensitive to the CP phase for $L=1000$ km
and $5000$ km. In order to quantify the CP phase dependence of
$P_{e\mu}$, we study peak values of $P_{e\mu}$, which occur at
energies described by the curve {\bf max1} in Fig.~\ref{maxmin} for
different baseline lengths. This peak value for a specific baseline
length depends on the CP violation phase $\delta_{\rm CP}$ and we
denote the maximum and the minimum of this value as $P_{e\mu}^{\rm
max}$ and $P_{e\mu}^{\rm min}$ respectively. The difference and the
ratio of these two values as functions of the baseline length $L$
are shown in Fig.~\ref{emu-cp}.
It is interesting to note that the ratio $P_{e\mu}^{\rm
min}/P_{e\mu}^{\rm max}$ increases monotonically with the baseline
length until $L=7600$ km. The ratio begins to decrease for a larger
baseline but remains larger than $90\%$. In fact, one can see that
$P_{e\mu}^{\rm max}$ and $P_{e\mu}^{\rm min}$ differ by less than
$10\%$ for $L\geq 6500$ km. We point out that $1-P_{e\mu}^{\rm
min}/P_{e\mu}^{\rm max}$ reaching to minimum at $L=7600$ km confirms
the so-called magic baseline for the probability $P_{e\mu}$
\cite{Barger:2001yr,Huber:2003ak,Smirnov:2006sm}.

\subsection{The dependencies of $P_{\mu\mu}$ and $P_{e\mu}$ on mixing angles $\theta_{13}$ and $\theta_{23}$}
Having studied CP phase dependencies of oscillation probabilities,
we now focus on $\theta_{13}$ and $\theta_{23}$ dependencies. In
this study we set the CP phase $\delta_{\rm CP}$ equal to zero.
\begin{figure}[tb]
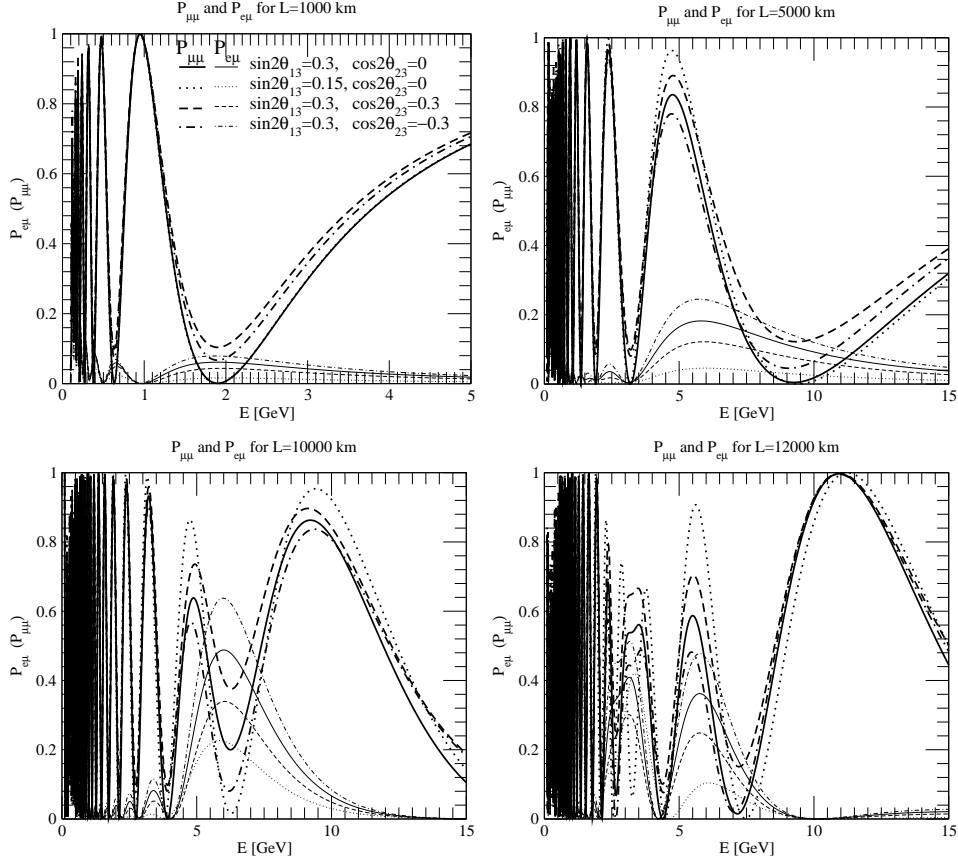

\begin{center}
$\begin{array}{cc}
\includegraphics*[width=6.2cm]{FL1000_sc.eps} &
\includegraphics*[width=6.2cm]{FL5000_sc.eps}
\\
\includegraphics*[width=6.2cm]{FL10000_sc.eps} &
\includegraphics*[width=6.2cm]{FL12000_sc.eps}
\end{array}$
\end{center}
\caption{The $\theta_{13}$ and $\theta_{23}$ dependencies of
$P_{\mu\mu}$ and $P_{e\mu}$ for $L=1000$ km, $5000$ km, $10000$ km
and $12000$ km respectively. The probability $P_{\mu\mu}$ is
described by the thick curve while $P_{e\mu}$ is described by the
thin curve. } \label{theta1323}
\end{figure}
The results are presented in Fig.~\ref{theta1323}.  It is easily
seen that the values of $P_{\mu\mu}$ at its local maximum and local
minimum depend on mixing angles $\theta_{13}$ and $\theta_{23}$
while only the local maximum of $P_{e\mu}$ depends on these
parameters. This confirms our earlier comments concerning the left
panel of Fig.~\ref{maxmin}. We point out that the differences
between solid and dotted curves in Fig.~\ref{theta1323} reflect the
effect of $\sin2\theta_{13}$; while the differences between solid
and dashed curves there reflect the effect of $\cos2\theta_{23}$.

\begin{figure}[tb]
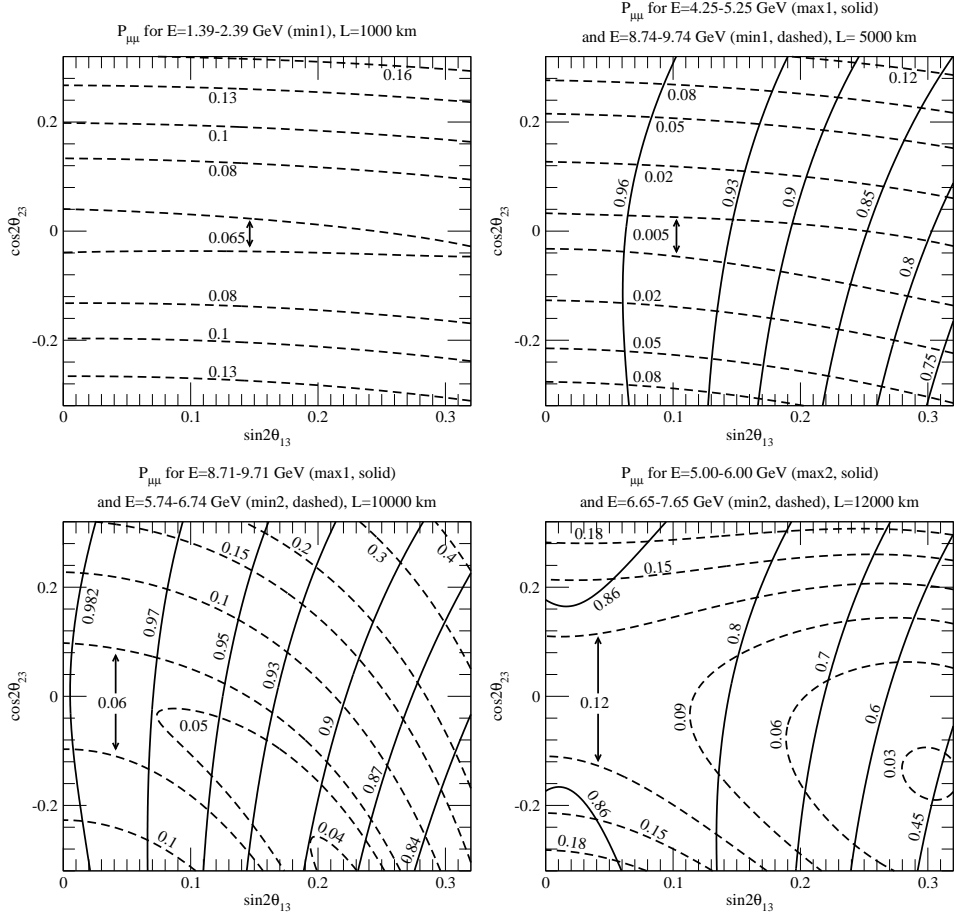

\begin{center}
$\begin{array}{cc}
\includegraphics*[width=6.2cm]{Fcontour_pmmm_L1000.eps} &
\includegraphics*[width=6.2cm]{Fcontour_pmmm_L5000.eps}
\\
\includegraphics*[width=6.2cm]{Fcontour_pmmm_L10000.eps} &
\includegraphics*[width=6.2cm]{Fcontour_pmmm_L12000.eps}
\end{array}$
\end{center}
\caption{The contour graphs of the muon neutrino survival
probability $P_{\mu\mu}$ on the  $\cos 2\theta_{23}-\sin
2\theta_{13}$ plane. At $L=1000$ km, the local minimum of
$P_{\mu\mu}$ on the curve {\bf min1} occurs at $E=1.89$ GeV. We plot
the contour graph of $P_{\mu\mu}$ by averaging this probability over
an $1$ GeV energy range centered at the above local minimum. At
$L=5000$ km, the local maximum of $P_{\mu\mu}$ on the curve {\bf
max1} occurs at $E=4.75$ GeV while the local minimum of this
probability on the curve {\bf min1} occurs at $E=9.24$ GeV. We plot
the contour graphs of $P_{\mu\mu}$ in the energy range $4.25\leq
E/{\rm GeV}\leq 5.25$ for the former case and $8.74\leq E/{\rm
GeV}\leq 9.74$. The same type of convention applies to $L=10000$ km
and $12000$ km.} \label{contour-pmm}
\end{figure}
We now present contour graphs of $P_{\mu\mu}$ and $P_{e\mu}$ on the
$\cos 2\theta_{23}-\sin 2\theta_{13}$ plane. The range for
$\cos2\theta_{23}$ is chosen such that $\sin2\theta_{23}>0.9$
\cite{Ashie:2004mr}, i.e., $-0.316<\cos2\theta_{23}<0.316$; while
$\sin^2 2\theta_{13}$ is chosen to be less than $0.1$, i.e.,
$\sin2\theta_{13}<0.316$. The contour graphs of $P_{\mu\mu}$ at
different baseline lengths are presented in Fig.~\ref{contour-pmm}.
Except for $L=1000$ km, we have shown contours of $P_{\mu\mu}$ for
energies in the vicinity of both local maximum and local minimum of
this probability. The contour for the local maximum of $P_{\mu\mu}$
at $L=1000$ km is not shown since $P_{\mu\mu}$ at this energy and
baseline length is not sensitive to mixing angles $\theta_{13}$ and
$\theta_{23}$. For $L=5000$ km, $10000$ km and $12000$ km, it is
seen that the contours at local maxima of $P_{\mu\mu}$ and those at
local minima of $P_{\mu\mu}$ behave rather differently. The former
are in general more parallel to the $\cos2\theta_{23}$-axis while
the latter are generally more parallel to the
$\sin2\theta_{13}$-axis. We note that the local maximum ({\bf max2})
of $P_{\mu\mu}$ at $L=12000$ km can vary from $0.9$ to a much
smaller value, 0.45, which is a result of significant matter
effects. Similarly, due to large matter effects, the local minimum
({\bf min2}) of $P_{\mu\mu}$ at $L=10000$ km can vary from $0$ to a
much larger value, $0.4$. We also notice that, at this baseline
length, the $\theta_{23}$ degeneracy is absent for $P_{\mu\mu}>0.1$.
In general, such a degeneracy is also absent for energies near local
maxima of $P_{\mu\mu}$. However, the probabilities are not sensitive
to $\cos2\theta_{23}$ in those cases. For comparisons, we also
present contour graphs for the appearance probability $P_{e\mu}$ at
different baseline lengths.
\begin{figure}[tb]
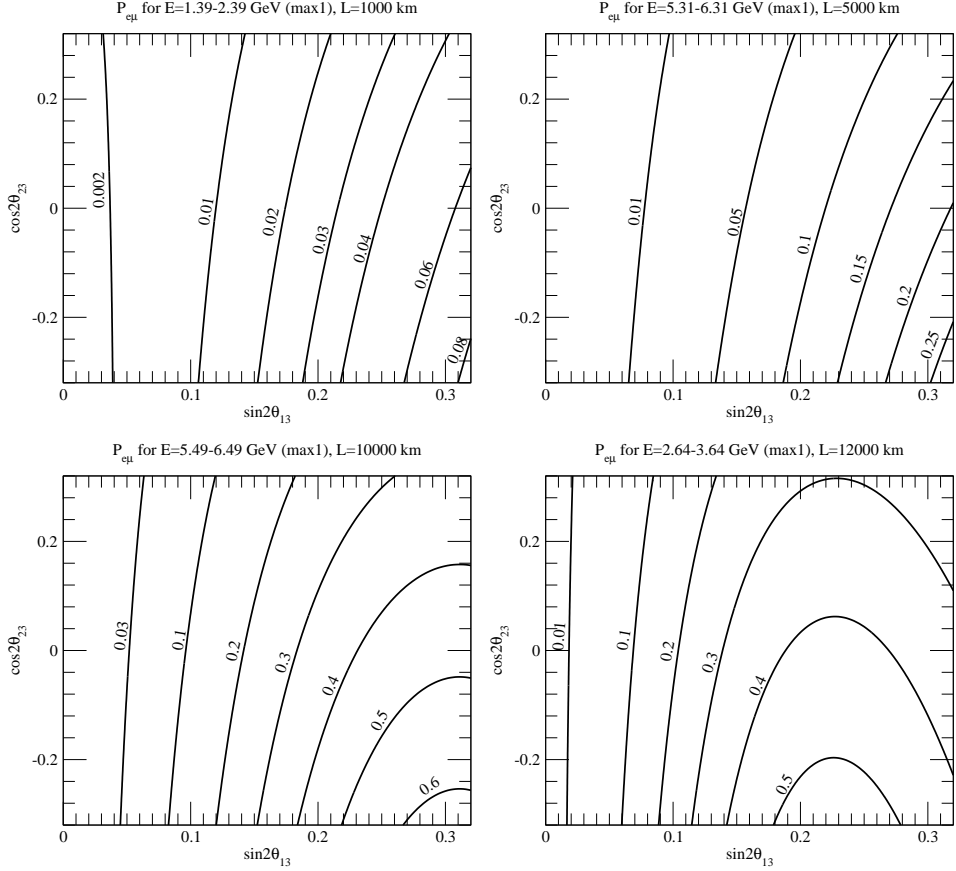

\begin{center}
$\begin{array}{cc}
\includegraphics*[width=6.2cm]{Fcontour_pmme_L1000.eps} &
\includegraphics*[width=6.2cm]{Fcontour_pmme_L5000.eps}
\\
\includegraphics*[width=6.2cm]{Fcontour_pmme_L10000.eps} &
\includegraphics*[width=6.2cm]{Fcontour_pmme_L12000.eps}
\end{array}$
\end{center}
\caption{The contour graphs for the oscillation probability
$P_{e\mu}$ on the $\cos 2\theta_{23}-\sin 2\theta_{13}$ plane. We
plot contours of $P_{e\mu}$ at energies near the local maximum ({\bf
max1}) of this probability. } \label{contour-pem}
\end{figure}
It is clearly seen that $P_{e\mu}$ is only sensitive to
$\sin2\theta_{13}$ for most cases. The sensitivity to
$\cos2\theta_{23}$ only occurs at very long baseline lengths and
large values of $\sin2\theta_{13}$. For example, at $L=10000$ km,
$P_{e\mu}$ becomes sensitive to $\cos2\theta_{23}$ as
$\sin2\theta_{13}$ approaches $0.3$. At $L=12000$ km, $P_{e\mu}$
becomes sensitive to $\cos2\theta_{23}$ when $\sin2\theta_{13}$ is
greater than $0.2$.

\subsection{A global look at the absence of $\theta_{23}$ degeneracy}
In this subsection, we focus on $\theta_{23}$ dependencies of
$P_{\mu\mu}$ and $P_{e\mu}$ for general baseline lengths. The
two-layer analytic approximations for $P_{\mu\mu}$ and $P_{e\mu}$
will  be employed for our discussions, and observations in the
previous subsection shall be justified. It is instructive to rewrite
Eq.~(\ref{analytic}) in polynomials of $\cos2\theta_{23}$:
\begin{eqnarray}
f(y,z)&=&-\alpha  y^2+(\alpha+\beta)y+(1-\beta), \nonumber \\
g(y,z)&=&-\gamma (y-1),
 \label{quadratic}
\end{eqnarray}
with $f(y,z)\equiv P_{\mu\mu}$, $g(y,z)\equiv P_{e\mu}$, $y\equiv
\cos 2\theta_{23}$ and $z\equiv \sin 2\theta_{13}$. Furthermore,
\begin{eqnarray}
\alpha&=&-\frac{1}{4}\left[\left(u-\cos t\right)^2+\left(v-\sin
t\right)^2\right],\nonumber \\
\beta&=&\frac{1}{2}\left(1-u^2-v^2\right)+\frac{1}{4}\left[\left(u-\cos
t\right)^2+\left(v-\sin t\right)^2\right],\nonumber \\
\gamma &=&\frac{1}{2}\left(1-u^2-v^2\right).
\end{eqnarray}
We note that the $\sin2\theta_{13}$ dependencies of $P_{\mu\mu}$ and
$P_{e\mu}$ reside in quantities $u$, $v$, $\cos t$ and $\sin t$.
These quantities also depend on the baseline length $L$ and the
neutrino energy $E$. Hence the coefficients $\alpha$, $\beta$ and
$\gamma$ also depend on the baseline length $L$ and the neutrino
energy $E$. It is interesting to note that $\alpha+\beta=\gamma$.
Therefore we have
\begin{equation}
P_{\mu\tau}=\alpha\left(y^2-1\right),
\end{equation}
using $P_{\mu e}+P_{\mu\mu}+P_{\mu\tau}=1$ and $P_{\mu e}=P_{e\mu}$
with our choice of $\delta_{\rm CP}=0$.
\begin{figure}[tb]
\begin{center}
$\begin{array}{c}
\includegraphics[width=9.5cm]{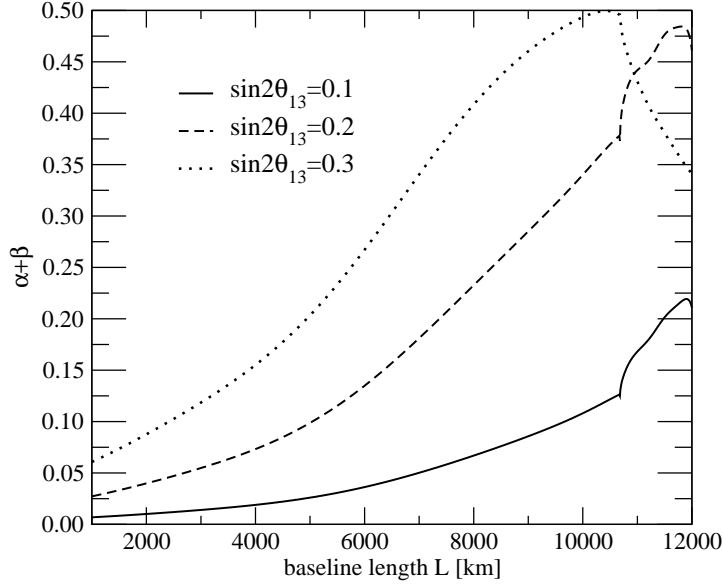}
\end{array}$
\end{center}
\caption{The coefficient $\alpha(z)+\beta(z)$ calculated along the
energy curve {\bf max1} in the left panel of Fig.~\ref{maxmin}. The
values of $\sin2\theta_{13}$ are taken to be $0.1, \ 0.2$ and $0.3$
respectively. } \label{pem}
\end{figure}
The contour structure of $P_{e\mu}$ is straightforward as $g(y,z)$
is only a linear function of $y$. Hence no $\theta_{23}$ degeneracy
presents in the contour graphs depicted in Fig.~\ref{contour-pem}.
Additionally, the sensitivity of $P_{e\mu}$ to $\cos2\theta_{23}$ is
$dg(y,z)/dy=-(\alpha(z)+\beta(z))$. The coefficient $\alpha+\beta$
evaluated along the energy curve {\bf max1} in the left panel of
Fig.~\ref{maxmin} are plotted in Fig.~\ref{pem} for
$\sin2\theta_{13}=0.1, \ 0.2$ and $0.3$. For $\sin2\theta_{13}=0.3$,
$\alpha+\beta$ reaches to the maximal value, 0.5, for $L\simeq
10500$ km. For $\sin2\theta_{13}=0.1$ and $0.2$, $\alpha+\beta$
rises quickly as the baseline length $L$ surpasses $10674$ km. We
note that the value of $P_{e\mu}$ is proportional to $\alpha+\beta$.
Hence $\alpha+\beta$ shown in Fig.~\ref{pem} is its own maximal
value for each baseline length $L$.

The contour structure of $P_{\mu\mu}$ can be analyzed through the
quadratic polynomial $f(y,z)$ in $y$. If $-\alpha\gg \alpha+\beta$,
generally there are two solution curves for $f(y,z)=p$ compatible
with the ranges of $y$ and $z$ where $p$ is a given value for
$P_{\mu\mu}$. Let us suppose that $z\equiv \sin2\theta_{13}$ is
measured in the future \cite{Ardellier:2006mn,Wang:2006ca} with a
central value $z_0$. The two solution curves for the equation
$f(y,z)=p$ then intersect with the straight line $z\equiv
\sin2\theta_{13}=z_0$ at two points $(y_1,z_0)$ and $(y_2,z_0)$.
This is actually what we have seen in Fig.~\ref{contour-pmm} for
local minima of $P_{\mu\mu}$. If, on the other hand, $-\alpha\simeq
\alpha+\beta$ or even $-\alpha\ll \alpha+\beta$, there exists only
one solution curve for the equation $f(y,z)=p$. This is because that
the two points $(y_1,z_0)$ and $(y_2,z_0)$ can not simultaneously
satisfy the constraint $-0.316<y<0.316$ since $\vert
y_1+y_2\vert=-(\alpha+\beta)/\alpha\geq 1$. This is actually what we
have seen in Fig.~\ref{contour-pmm} for local maxima of
$P_{\mu\mu}$. To justify this observation, it remains to show that
the coefficient $-\alpha$ dominates over $\alpha+\beta$ at energies
corresponding to local minima of $P_{\mu\mu}$ while the latter
dominates over the former at energies corresponding to local maxima
of $P_{\mu\mu}$.
\begin{figure}[tb]
\begin{center}
$\begin{array}{cc}
\includegraphics[width=6.2cm]{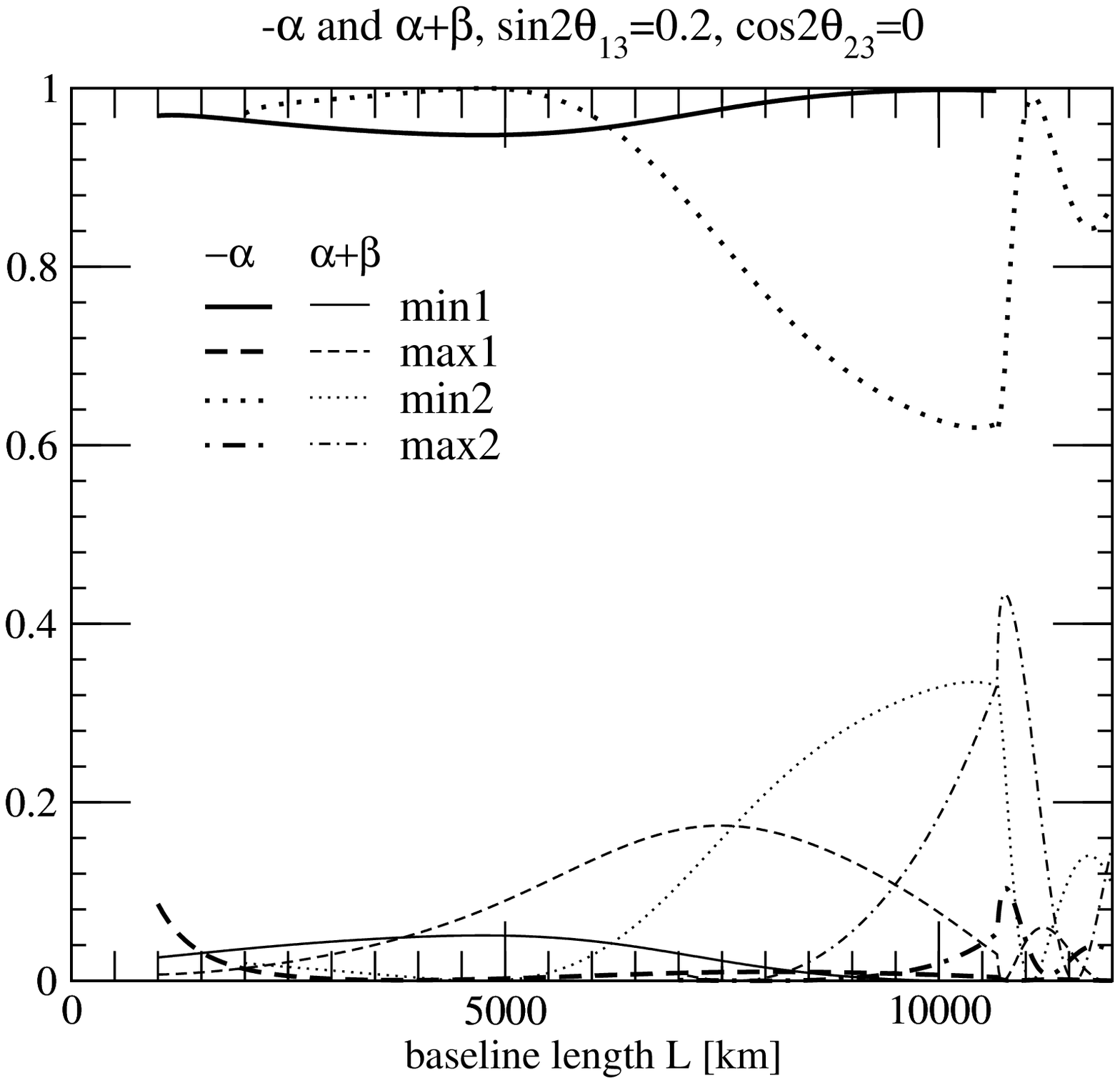} &
\hspace{1cm}
\includegraphics[width=6.2cm]{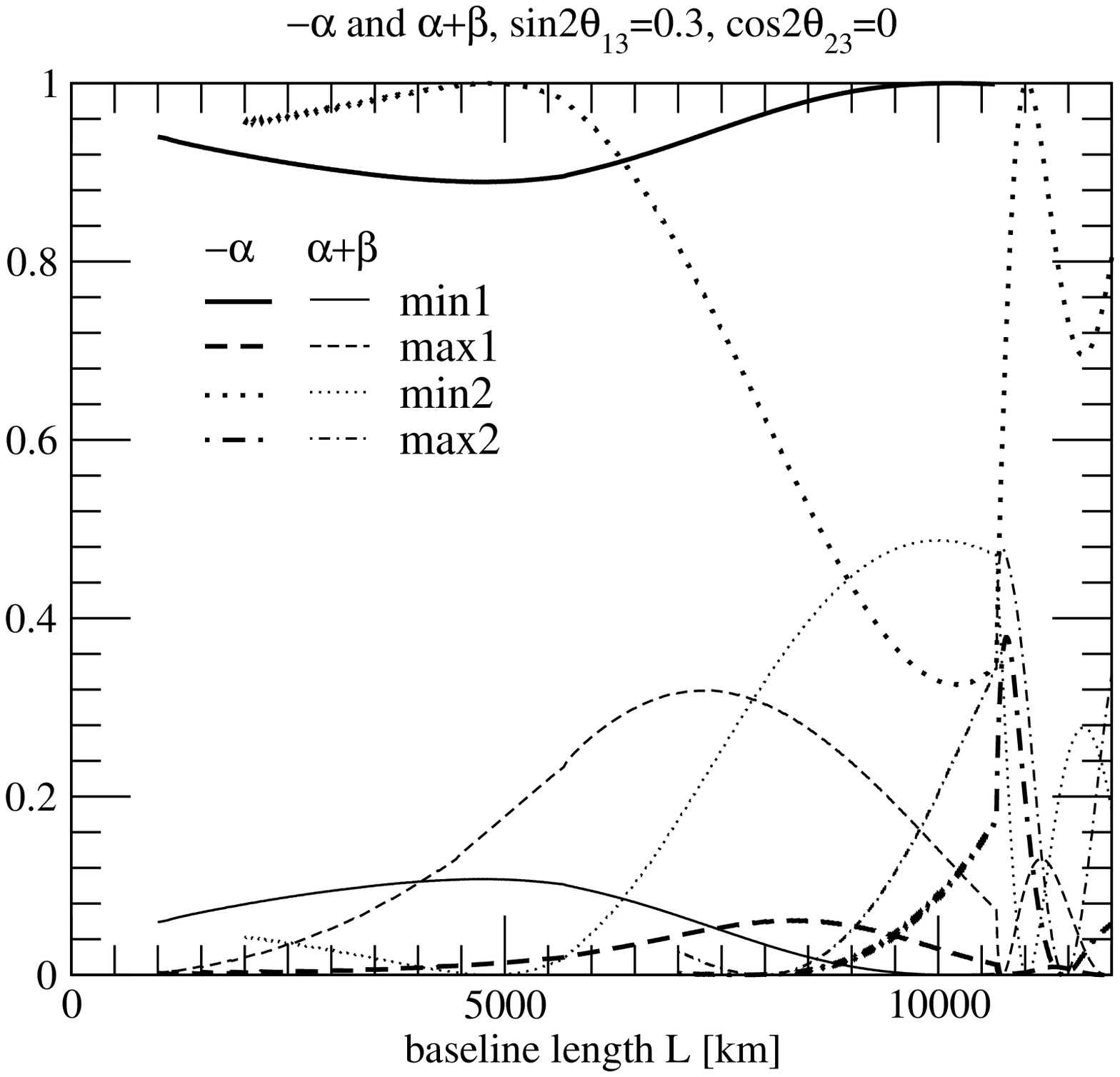}
\end{array}$
\end{center}
\caption{The coefficients $-\alpha$ and $\alpha+\beta$ evaluated
along energy curves in the right panel of Fig.~\ref{maxmin}. The
coefficients are calculated with $\sin2\theta_{13}=0.2$ and $0.3$ on
the left and right panels respectively. The thick curves denote
values of $-\alpha$ while thin curves denote those of
$\alpha+\beta$. For solid and dotted curves, the thick curves
generally dominate over the corresponding thin ones. For dashed and
dotted-dashed curves, the thin curves generally dominate over the
thick ones.} \label{coefficient}
\end{figure}
This is clearly demonstrated in Fig.~\ref{coefficient} where the
coefficients $-\alpha$ and $\alpha+\beta$ are evaluated along energy
curves in the right panel of Fig.~\ref{maxmin}. We have calculated
the coefficients with $\cos2\theta_{23}=0$ and
$\sin2\theta_{13}=0.2, \ 0.3$ respectively. We remark that other
choices for $\cos2\theta_{23}$ do not produce noticeable changes on
the energy curves where $-\alpha$ and $\alpha+\beta$ are evaluated.

It is easily seen that $\alpha+\beta$ always dominates over
$-\alpha$ when these coefficients are evaluated at energies along
{\bf max1} or {\bf max2} in the right panel of Fig.~\ref{maxmin}. In
such cases, the $\theta_{23}$ degeneracy is absent in the solutions
of $f(y,z)=p$. Namely there exists only one solution curve for the
above equation. Reversely, $-\alpha$ always dominates over
$\alpha+\beta$ when these coefficients are evaluated at energies
along {\bf min1}. The situation is slightly more complicated when
these coefficients are evaluated at energies along {\bf min2}. In
this case $-\alpha$ no longer dominates over $\alpha+\beta$ for
baseline lengths around $10^4$ km. In fact, with
$\sin2\theta_{13}=0.3$, $\alpha+\beta$ is even larger than $-\alpha$
for $9000\leq L/{\rm km}\leq 10500$. This explains the contour
structure of $P_{\mu\mu}$ at $L=10000$ km (see
Fig.~\ref{contour-pmm}) where the straight line $z=0.3$ only
intersects one equal probability curve $f(y,z=0.3)=p$. The straight
line $z=0.2$ also behaves the same except for a very small $p$. We
reiterate that the range for $y\equiv \cos2\theta_{23}$ is $-0.316<
y< 0.316$ due to the constraint $\sin^2 2\theta_{23}>0.9$
\cite{Ashie:2004mr}. Therefore, given $z=z_0$, the equation
$f(y,z_0)=p$ could have only one solution for $y$ if
$-(\alpha(z_0)+\beta(z_0))/\alpha(z_0) > 0.632$. In other words,
such values of $-(\alpha(z_0)+\beta(z_0))/\alpha(z_0)$ lead to the
absence of $\theta_{23}$ degeneracy. In fact, the condition for the
absence of $\theta_{23}$ degeneracy is even more relaxed. To see
this, let us divide our discussions according to the true octant of
$\theta_{23}$.
\subsubsection{ {\bf min2}, $\theta_{23}<\pi/4$}
Since $\alpha<0$ and $\alpha+\beta>0$, the two solutions for $y$ in
$f(y,z_0)=p$ are both negative for $1-\beta(z_0)-p>0$ while they
have opposite signs for $1-\beta(z_0)-p<0$. If the true value of
$\theta_{23}$ is less than $\pi/4$, i.e., the true value of $y$ is
positive, then the experimental measurement should give
$1-\beta(z_0)-p<0$ so that a positive solution for $y$ exists. With
$1-\beta(z_0)-p<0$, the two solutions for $f(y,z_0)=p$ have opposite
signs and the negative solution has a larger absolute value. The
negative solution will violate the constraint $-0.316<y$ if
$-(\alpha(z_0)+\beta(z_0))/\alpha(z_0)
> 0.316$.  For $z_0=0.2$,
$-(\alpha(z_0)+\beta(z_0))/\alpha(z_0)
> 0.316$ is valid for $8300\leq L/{\rm km}\leq 10770$. For $z=0.3$, the
above baseline range is extended to $7410\leq L/{\rm km}\leq 10790$.
\subsubsection{ {\bf min2}, $\theta_{23}>\pi/4$}
With a true value of $\theta_{23}$ greater than $\pi/4$, i.e., the
true value of $y$ less than zero, the value of $1-\beta(z_0)-p$ can
either be positive or negative. The condition
$1-\beta(z_0)-p>(<)\,0$ is equivalent to the condition
$\left|y\right|<(>)\left(\alpha(z_0)+\beta(z_0)\right)/\left(-\alpha(z_0)\right)$.
For $-(\alpha(z_0)+\beta(z_0))/\alpha(z_0)
> 0.316$, one must have $1-\beta(z_0)-p>0$. Hence there exist two
negative solutions for $y$. In this case, the corresponding
solutions for $\theta_{23}$ are both located in the same octant. For
$0.316< -(\alpha(z_0)+\beta(z_0))/\alpha(z_0)<0.632$, the spurious
solution for $y$ may or may not violate the constraint $y>-0.316$.
For $-(\alpha(z_0)+\beta(z_0))/\alpha(z_0)
> 0.632$, the spurious solution for $y$ must violate the constraint
$y>-0.316$, hence the $\theta_{23}$ degeneracy is surely absent. For
$\sin2\theta_{13}=0.2$, the condition
$-(\alpha(z_0)+\beta(z_0))/\alpha(z_0) > 0.632$ can not be achieved
along {\bf min2}. For $\sin2\theta_{13}=0.3$, the above condition is
satisfied for $8270\leq L/{\rm km}\leq 10720$. For
$-(\alpha(z_0)+\beta(z_0))/\alpha(z_0< 0.316$, $1-\beta(z_0)-p$ can
either be positive or negative. For $1-\beta(z_0)-p>0$, both
solutions for $y$ are negative and satisfying the constraint $y
> -0.316$. For $1-\beta(z_0)-p<0$, both solutions for $y$ satisfy the
constraint $-0.316<y<0.316$. However their corresponding
$\theta_{23}$ angles are situated in different octants.

Let us summarize the results obtained in this subsection. The
coefficient $\alpha+\beta$ dominates over $-\alpha$ for energy
values along curves {\bf max1} and {\bf max2} for all baseline
lengths. Hence the $\theta_{23}$ degeneracy is absent along these
energy curves for all baseline lengths. The situation along the
curve {\bf min1} is just the opposite, the coefficient $-\alpha$
dominates over $\alpha+\beta$ for all baseline lengths. Hence the
$\theta_{23}$ degeneracy is present for all baseline lengths in this
case. The issue of $\theta_{23}$ degeneracy becomes more complicated
along {\bf min2}, which we have discussed according to the true
octant of $\theta_{23}$. Along the energy curve {\bf min2}, the
non-degeneracy baseline range is larger for the $\theta_{23}<\pi/4$
case.

\section{Discussions and Conclusions}
We have presented the baselines and energies ideal for probing the
octant of $\theta_{23}$ through neutrino oscillations. The
appearance mode $\nu_e\to \nu_{\mu}$ can be studied in a very long
baseline with the facility of neutrino factory \cite{nu_factory} or
the more recent proposed $\beta$ beam \cite{Zucchelli:2002sa}.  As
said, the sensitivity of $P_{e\mu}$ to $\theta_{23}$ is
$dg(y,z)/dy=-(\alpha(z)+\beta(z))$ where $g(y,z)$ represents
$P_{e\mu}$ in the analytic approximation given by
Eq.~(\ref{quadratic}). The maximal value of $\alpha+\beta$ for each
baseline length is shown in Fig.~\ref{pem}. At the magic baseline,
$L=7600$ km, $\alpha+\beta=0.06, \ 0.21$ and $0.38$ for
$\sin2\theta_{13}=0.1, \ 0.2$ and $0.3$ respectively. For a
sufficiently large $\sin2\theta_{13}$ and a baseline length close to
the magic value \cite{Barger:2001yr,Huber:2003ak,Smirnov:2006sm},
$P_{e\mu}$ is ideal for probing the octant of $\theta_{23}$.

The probability $P_{\mu\mu}$ is also relevant in neutrino
oscillation experiments with neutrino factories. The sensitivity of
this probability to $\theta_{23}$ is determined by the derivative
\begin{equation}
r\equiv \frac{dP_{\mu\mu}}{d\cos2\theta_{23}}=-2\alpha
\cos2\theta_{23}+\left(\alpha+\beta\right).
\label{sensitivity}
\end{equation}
Since $\alpha$ is negative, the sensitivity $r$ is larger for $\cos
2\theta_{23}>0$, i.e., $\theta_{23}<\pi/4$. For a measurement
performed around a local maximum of $P_{\mu\mu}$, the sensitivity to
$\theta_{23}$ is completely determined by the coefficient
$\alpha+\beta$, since the coefficient $\alpha$ is generally rather
suppressed in this case. Along the energy curve denoted by {\bf
max1}, $\alpha+\beta$ peaks at $7480$ km for $\sin2\theta_{13}=0.2$
and it peaks at $L=7350$ km for $\sin2\theta_{13}=0.3$. The values
of $\alpha+\beta$ at those peaks are $0.17$ and $0.32$ respectively.
Along the curve {\bf max2}, $\alpha+\beta$ peaks around $L=10750$ km
for both $\sin 2\theta_{13}=0.2$ and $0.3$ with values  $0.43$ and
$0.48$ respectively.

For a measurement performed around a local minimum of $P_{\mu\mu}$,
the sensitivity to $\theta_{23}$ is determined by both coefficients
$-\alpha$ and $\alpha+\beta$. Along the energy curve denoted by {\bf
min1}, the coefficient $-\alpha$ is always close to unity while the
coefficient $\alpha+\beta$ is always suppressed for all baseline
lengths. It is understood that the magnitude of $\alpha+\beta$
determines the size of matter effects. Hence the matter effect is
small at energies along the curve {\bf min1}. The suppression of
$\alpha+\beta$ compared to $-\alpha$ leads to the $\theta_{23}$
degeneracy as discussed before. The behavior of $P_{\mu\mu}$ along
the energy curve {\bf min2} is more interesting. If the true value
of $\theta_{23}$ is less than $\pi/4$, the $\theta_{23}$ degeneracy
from the measurement of $P_{\mu\mu}$ is absent in the baseline range
$8300\leq L/{\rm km}\leq 10770$ for $\sin 2\theta_{13}=0.2$. The
above non-degeneracy baseline range extends to $7410\leq L/{\rm
km}\leq 10790$ for $\sin 2\theta_{13}=0.3$. On the other hand, if
the true $\theta_{23}$ is greater than $\pi/4$, the non-degeneracy
baseline range does not exist along the energy curve {\bf min2} for
$\sin 2\theta_{13}=0.2$. For $\sin 2\theta_{13}=0.3$, the
non-degeneracy baseline range is $8270\leq L/{\rm km}\leq 10720$.

The existence of non-degeneracy baseline range along the energy
curve {\bf min2} has important implications. It can be seen from
Fig.~\ref{maxmin} that the curve {\bf min2} lies in between curves
{\bf max1} and {\bf max2}. Since the degeneracy of $\theta_{23}$ is
absent on both {\bf max1} and {\bf max2} for all baselines, it is
possible that there exists a non-degeneracy region spanned by ranges
of the baseline length and the neutrino energy. For example, with
$\sin2\theta_{13}=0.2$ and a true value of $\theta_{23}$ less than
$\pi/4$, the $\theta_{23}$ degeneracy is absent for $8300\leq L/{\rm
km}\leq 10770$ for energies along curves {\bf max2}, {\bf min2} and
{\bf max1}. It is of great interest to investigate if the
$\theta_{23}$ degeneracy is also absent for any neutrino energy
larger than the value on {\bf max2} and smaller than that on {\bf
max1}. By taking all these energies into account, we find that the
$\theta_{23}$ degeneracy is absent for $8550\leq L/{\rm km}\leq
10680$. For a true value of $\theta_{23}$ greater than $\pi/4$ and
$\sin2\theta_{13}=0.3$, the $\theta_{23}$ degeneracy is absent for
$8270\leq L/{\rm km}\leq 10720$ for energies along curves {\bf
max2}, {\bf min2} and {\bf max1}. However, with all energies between
curves {\bf max2} and {\bf max1} considered, we find that the
$\theta_{23}$ degeneracy is absent for $8450\leq L/{\rm km}\leq
10680$. The non-degeneracy baseline range corresponding to different
combinations of $\theta_{23}$ and $\theta_{13}$ values are
summarized in Table~I.
\begin{table}[bt]
\begin{center}
\caption{The baseline range in which the $\theta_{23}$ degeneracy is
absent in the probability $P_{\mu\mu}$ for all energy values between
the curves {\bf max2} and {\bf max1}. The entry corresponding to
$\theta_{23}>\pi/4$ and $\sin2\theta_{13}=0.2$ is left blank since,
with such set of parameters, there exists no baseline length where
the condition for the absence of $\theta_{23}$ degeneracy can be
satisfied. }
\begin{tabular}{|c|c|c|}\hline
$\theta_{23}$ octant & $\sin2\theta_{13}=0.2$ &
$\sin2\theta_{13}=0.3$ \\ \hline $\theta_{23}<\pi/4$ & $8550\leq
L/{\rm km}\leq 10680$ & $7950\leq L/{\rm km}\leq 10700$ \\
\cline{2-3} $\theta_{23}>\pi/4$ &  &$8450\leq L/{\rm km}\leq
10680$\\ \hline
\end{tabular}
\end{center}
\label{table}
\end{table}

It is interesting to compare measurements on $P_{e\mu}$ and
$P_{\mu\mu}$ since both oscillations appear in experiments with
neutrino factories \cite{nu_factory}. The main issue for comparison
is on the determination of the true $\theta_{23}$ value under the
assumption that both the sign of $\Delta m_{31}^2$ and the value of
$\sin2\theta_{13}$ are known. Let us begin the discussion with a
true value of $\theta_{23}$ less than $\pi/4$ and
$\sin2\theta_{13}=0.2$. For $L< 8550$ km, the appearance mode
$\nu_{e}\to \nu_{\mu}$ is useful for probing the octant of
$\theta_{23}$, in particular for $L$ close to the magic value,
$7600$ km. However, the survival mode $\nu_{\mu}\to \nu_{\mu}$ is
not as useful since the $\theta_{23}$ degeneracy is absent only at
energies near {\bf max1} and {\bf max2}. Concerning the sensitivity
to $\theta_{23}$, we note that the differentiation of $P_{e\mu}$
with respect to $\cos2\theta_{23}$ is $-(\alpha+\beta)$. The value
of $\alpha+\beta$ increases with $L$ as shown in Fig.~\ref{pem}. It
is $0.21$ at $L=7600$ km, and $0.26$ at $L=8550$ km. For $8550\leq
L/{\rm km}\leq 10680$, both $\nu_e\to \nu_{\mu}$ and $\nu_{\mu}\to
\nu_{\mu}$ are useful for probing the octant of $\theta_{23}$. We
note that peak positions of $P_{e\mu}$ are mostly around $6$ GeV.
Hence they overlap with the non-degeneracy energy range of
$P_{\mu\mu}$. From Eq.~(\ref{sensitivity}) and our assumption of
$\theta_{23}<\pi/4$, we find that $P_{\mu\mu}$ is more sensitive to
$\theta_{23}$ as compared to $P_{e\mu}$ for the same neutrino
energy. For $L>10680$ km, $\nu_e\to \nu_{\mu}$, is again the only
useful mode for probing the octant of $\theta_{23}$.

Let us turn to the case where the true value of $\theta_{23}$ is
greater than $\pi/4$ and $\sin2\theta_{13}=0.2$. In such a case, the
value for $P_{e\mu}$ is enhanced compared to that with $y>0$.
Furthermore $P_{e\mu}$ is always more sensitive to
$\cos2\theta_{23}$ as compared to $P_{\mu\mu}$ for the same neutrino
energy. It is clear that the $\nu_e\to \nu_{\mu}$ appearance mode is
more useful for probing $\theta_{23}$ regardless the baseline
length. Although there exists a baseline range where the
$\theta_{23}$ degeneracy is absent in $P_{\mu\mu}$ for neutrino
energies between curves {\bf max2} and {\bf max1}. However this
requires a large value of $\sin2\theta_{13}$, such as
$\sin2\theta_{13}=0.3$.

It is essential to remark that the above non-degeneracy baseline
range is not sensitive to the value of $\Delta m_{31}^2$, which we
have so far taken to be $2.4\cdot 10^{-3}$ eV$^2$. Changing the
value of $\Delta m_{31}^2$ only shifts the probability curves in
Fig.~\ref{cp-phase} and Fig.~\ref{theta1323} so that positions for
local maxima and local minima of these probabilities shift
accordingly. However, the maximal or minimal values of these
probabilities remain unchanged. In other words, although the energy
curves in Fig.~\ref{maxmin} are shifted, the coefficients $-\alpha$
and $\alpha+\beta$ plotted in Fig.~\ref{coefficient}, which combine
to form $P_{e\mu}$ and $P_{\mu\mu}$ (see Eq.~(\ref{quadratic})),
remain the same. The values of these coefficients as functions of
the baseline length $L$ then determine the non-degeneracy baseline
range.

In conclusion, we have studied the probabilities $P_{e\mu}$ and
$P_{\mu\mu}$ for very long baseline neutrino oscillations. We focus
on sensitivities of these probabilities to mixing angles
$\theta_{13}$, $\theta_{23}$ and the CP violation phase $\delta_{\rm
CP}$. Taking $\delta_{\rm CP}=0$ as an example, we presented contour
graphs of $P_{e\mu}$ and $P_{\mu\mu}$ in the
$\sin2\theta_{13}-\cos2\theta_{23}$ plane for baseline lengths
$L=1000$ km, $5000$ km, $10000$ km and $12000$ km. The energy values
chosen for such studies are in the vicinities of either local minima
or local maxima of neutrino oscillation probabilities. For each
baseline length, we have found that $P_{\mu\mu}$ is more sensitive
to $\sin 2\theta_{13}$ at energies around its local maxima while it
is more sensitive to $\cos 2\theta_{23}$ at energies around its
local minima. On the other hand, the appearance probability
$P_{e\mu}$ is sensitive to $\sin2\theta_{13}$ and $\cos2\theta_{23}$
only near its local maximum. Such findings have been applied to
probe the octant of mixing angle $\theta_{23}$ assuming that the
angle $\theta_{13}$ and the sign of $\Delta m_{31}^2$ are known. The
appearance probability $P_{e\mu}$ is non-degenerate in
$\theta_{23}$. The sensitivity of $P_{e\mu}$ to $\cos2\theta_{23}$
is studied for baseline lengths from $1000$ km to $12000$ km. We
also studied the sensitivity of $P_{\mu\mu}$ to $\cos2\theta_{23}$
for the same range of baseline length. We have identified the ranges
of neutrino energy and baseline lengths where the $\theta_{23}$
degeneracy is absent. We have pointed out that, for a true value of
$\theta_{23}$ less than $\pi/4$ and a baseline length between $8000$
and $10000$ km, the survival mode $\nu_{\mu}\to \nu_{\mu}$ is
equally good as the appearance mode $\nu_{e}\to \nu_{\mu}$ for
probing the octant of $\theta_{23}$.

\section*{Acknowledgements}
G.L.L likes to thank D. Indumathi for informative discussions. This
work is supported by National Science Council of Taiwan under the
grant number NSC 94-2112-M-009-026.

\end{document}